\input psbox.tex
\documentstyle{l-aa}
\begin{document}

   \thesaurus{03         
             (07.01.1;   
              07.07.1;   
              07.17.1;   
              17.01.1)   
             }

\title{Cosmological evolution and large scale structures of radio
       galaxies and quasars\protect\footnote[1]}

\author{K.T. Chy\.{z}y and S. Zi\c{e}ba}

\offprints{K.T. Chy\.{z}y\protect\\$^{\star}$~Tables 1 and 2 are only available
       in electronic form at the CDS via anonymous ftp 130.79.128.5}

\institute{Astronomical Observatory of the Jagiellonian University\\
              ul. Orla 171, 30-244 Krak\'{o}w, Poland}

\date{Received 10 October 1994 / Accepted 13 March 1995}

\maketitle
\begin{abstract}
The simple unification scheme of powerful radio galaxies and quasars, based
entirely on the orientation dependent effects, has been confronted with the
observed radio structures for 152 radio galaxies and 173 steep spectrum
quasars. Contrary to the scheme's prediction, the cosmological evolution of
geometrical parameters describing the large scale structure of these two
types of radio sources are different. Linear size, arm ratio asymmetry and
bending are all together stronger evolving with epoch for radio galaxies.
Moreover, linear size and bending are more closely correlated with radio
luminosity for radio galaxies than for quasars. This supports the supposition
that, even if these two AGN classes are intrinsically identical sets of
objects in deep interior regions, their large scale structures reveal rather
various host environmental conditions which can lead to various classes of
objects.

   \end{abstract}

\section{Introduction}

The large scale structures of extended quasars and radio galaxies can be used
as a test for radio galaxy-quasar unification schemes. If both the mentioned
categories of radio sources are intrinsically the same type of objects but
only appear different to an observer due to the various viewing directions,
according to the Barthel's (1989) hypothesis, then their radio structures are
expected to evolve with redshift in the same way. This suggestion was put
forward by Gopal-Krishna and Kulkarni (1992) on the grounds of radio linear
sizes attained by extended quasars and powerful radio galaxies. However,
Chy\.{z}y and Zi\c{e}ba (1993, hereafter CZ), came recently, on the base of
152 radio galaxies and 173 quasars, to a quite contrary view indicating
differences not only in the cosmological evolution of radio galaxy and quasar
linear sizes but also in their size dependence on radio luminosity. A
conclusion supporting this point of view was also reported by Singal (1993b)
who examined a large sample of 789 sources. Singal (1993a) also showed that
observed relative numbers and linear sizes of radio galaxies and quasars from
3CR sample are inconsistent with unification even invoking a cosmic evolution
in the opening angle of obscuring torus $\psi$. In recent work Gopal-Krishna
et al. (1994) tried to bring into agreement Singal's results with the unified
scenario by incorporating a misalignment between the radio axis and the axis
of the visibility cone defined by the optically-thick torus surrounding the
nuclear region. However, their approach seems to be not convincing enough
regarding the involvement of rather large misalignment angles without any
discussion of their distribution.

In this paper we present further investigation of cosmic evolution of radio
structures based upon the other geometrical parameters which describe the
observed structures. Apart from the simplest linear size parameter $L$ it is
possible to determine two independent parameters assessing the asymmetry
of the structure: the arm lengths ratio $Q$, defined as the ratio of the
distances of hot spots from the core; and the misalignment $M$, which
measures the apparent bending, and is defined as the ratio of the
displacement of the core from the source axis to the linear size (see also
Fig. 1 in Zi\c{e}ba and Chy\.{z}y 1991, hereafter ZC).

The asymmetry parameters $Q$ and $M$ can potentially be a powerful tool in
the consistency test for the orientation based unification scheme as,
according to it, their evo\-lutionary patterns should be the same for
radio galaxies and quasars. Contrary to the linear size, they are not
sensitive to the simple homological rescaling of the whole struc\-ture and
hence to the age or expansion velocity of the structure. In that case,
possibly revealed differences in asym\-metry evolution of radio galaxies and
quasars might give evidence in favor of even deeper physical differences
between these two AGN types of sources. We performed~such a quantitative
comparison of radio galaxy and quasar apparent asymmetry, evaluating the
dependence of the $Q$ and $M$ parameters on redshift $z$ and spectral radio
luminosity $P$ at 1.4 GHz.

We also discussed the importance of projection effects using observed $Q$,
$M$ and $L$ distributions for finding the best fits of the kinematical model,
consistent with unification scenario, separately for radio galaxies and
quasars. A similar orientation modeling of radio source properties was also
presented by Lister et al. (1994). Another approach to the problem of
asymmetry observed among extragalactic radio sources based on the detailed
brightness distribution was recently published by Ry\'{s} (1994).

\section{The samples}

The observational base for our discussion comprises two samples which were
described and used in our earlier papers (ZC, CZ). The radio galaxy sample
contains 152 triple, edge-brightened FRII powerful objects (Fanaroff and
Riley 1974), carefully selected from the GB/GB2 complete sample (Machalski
and Maslowski 1982) and the 3CR sample (Spinrad et al. 1985), based mainly on
the compilation of Macklin (1981). The main
contributions to the quasar sample,
containing 173 objects, come from Barthel et al. list (1988), Hintzen et al.
(1983), Miley and Hartsuijker (1978), 3CR sample (Spinrad et al. 1985) and
GB/GB2 sources (Machalski and Maslowski 1982). To avoid undesirable bias all
possible subgalactic compact steep spectrum sources (linear size less than
10 kpc) had been extracted from the final sample.

Arm lengths ratio, misalignment and linear size were calculated from the
positions of the hot spots and the central component, which are usually found
in publications, or were estimated directly from maps. Radio flux densities
at 1.4 GHz and 4.85 GHz were taken from White and Becker (1992) and Becker et
al. (1991) catalogues respectively. Spectral indices and luminosities at 1.4
GHz in the emitted frame of the sources were derived from those data using an
Einstein-de Sitter Universe ($q_{0}=0.5$) with a Hubble constant
$H_{0}=100$ km s$^{-1}$ Mpc$^{-1}$. Table 1 and 2 (accessible in electronic
form) list all selected radio galaxies and quasars along with redshift,
luminosity, estimated geometrical parameters $Q$, $M$, $L$ and literature
data. The radio galaxy sample spans the redshift range $0.03<z<1.8$ and
the luminosity $10^{24.2}<P $[W Hz$^{-1}]<10^{28.1}$ and the quasar sample
spans respectively $0.1<z<2.7$ and $10^{25.1}<P $[W Hz$^{-1}]<10^{28.3}$.

\section{Results}

\begin{figure}
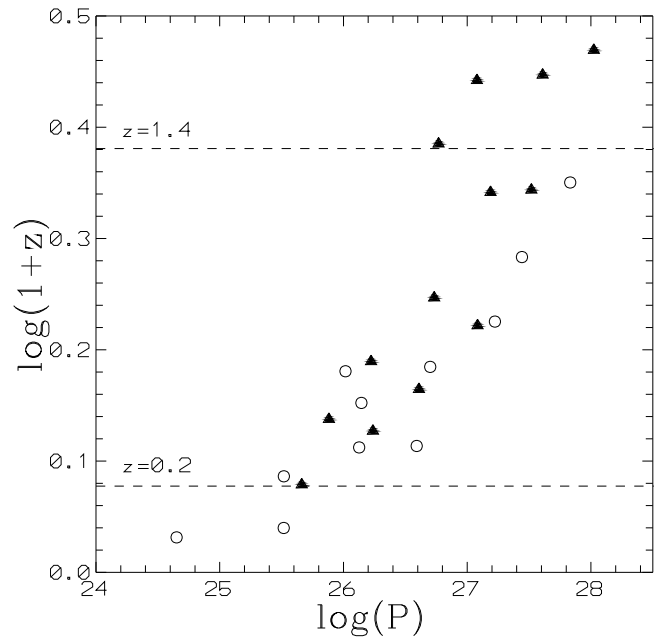

       \psbox{fig1a.psu}
      \caption{Median values of redshift versus luminosity for radio
       galaxy ($\circ$) and quasar ($\triangle$) groups resulting from the
       binning}
      \label{FigPz}
   \end{figure}

In order to derive the evolutionary behavior of the asymmetry of radio
structures we estimated the dependence of the median values of $Q$ and $M$
parameter on redshift and radio luminosity in the form: ${\propto}(1+z)^{n}$
and ${\propto}P^{\beta }$ respectively. As the coverage of the $P$-$z$ plane
by our radio galaxies and quasars is not uniform the special method was applied
to eliminate the influence of the observed redshift-luminosity correlation.
This method follows Oort's (1987) approach and was in detail described in our
earlier~paper (ZC). The first step consists in binning the data in broad
regions in the $P$-$z$ diagram, so that for equal radio~lumino\-sity bins (0.6
and 0.45 for radio galaxies and quasars respectively), the number of sources
in each two dimensional region $P$-$z$ was about $13(\pm3)$. The observed
median~values calculated for the resulting regions are presented in
Fig.~\ref{FigPz}.\par \vspace*{0pt}
\indent
The dependence of a median value of an asymmetry parameter on
redshift was fitted in subsequent steps, simultaneously with rescaling the
parameter to the chosen luminosity using the updated values of the
$P^{\beta }$ relation found in the previous step. The best solutions we have
found for the fitted $n$ and $\beta $ when this  approach was used on our
quasar and radio galaxy samples are listed in Table~\ref{QML} together with
95\% confidence intervals and analogous values estimated for linear sizes.
All numbers presented in Table~\ref{QML} were calculated in a uniform manner
using log$(Q, M, L)$-log$P$ and -log$(1+z)$ planes for rescaling of the
source parameters to a fixed radio luminosity $P=10^{26}$ W Hz$^{-1}$ and
redshift $z=0$. The radio galaxies were taken on the whole without dividing
them into classes of weak and bright objects as we did before (ZC,CZ). As
a consequence some numbers now obtained are slightly different from those
published earlier indicating this way how rescaling procedure influences
the results, however the obtained general picture is the same.
\begin{table}[t]
  \caption{The best-fit model parameters of asymmetry and linear size
  evolution of radio galaxies and quasars and 95\% confidence interval in
  parenthesis.}
  \label{QML}
\begin{center}
\begin{tabular}{ccc}
\hline
   &     $\beta$           &      $n$\\
\hline
\multicolumn{1}{l}{Radio Galaxies}\\
$Q$  &     $+0.05$ $\pm$0.01   & $-0.54$ $\pm$0.11\\
     &     $(+0.03,+0.07)$     & $(-0.79,-0.29)$ \\[2mm]
$M$  &     $-0.28$ $\pm$0.07 & $+2.91$ $\pm$0.62\\
     &     $(-0.44,-0.12)$     & $(+1.51,+4.31)$ \\[2mm]
$L$  &     $+0.35$ $\pm$0.03 & $-3.36$ $\pm$0.40\\
     &     $(+0.28,+0.42)$     & $(-4.26,-2.46)$ \\
\multicolumn{1}{l}{Quasars}\\
$Q$  &     $+0.01$ $\pm$0.01 & $-0.13$ $\pm$0.06\\
     &     $(-0.01,+0.03)$     & $(-0.26,+0.00)$ \\[2mm]
$M$  &     $+0.01$ $\pm$0.05 & $+0.28$ $\pm$0.27\\
     &     $(-0.10,+0.12)$     & $(-0.31,+0.87)$ \\[2mm]
$L$  &     $-0.10$ $\pm$0.05 & $-1.31$ $\pm$0.28\\
     &     $(-0.20,+0.01)$     & $(-1.93,-0.69)$ \\
\hline
 \end{tabular}
 \end{center}
\end{table}
The striking result of the comparison of the fitted parameters shown in
Table~\ref{QML} is the stronger evolution of the asymmetry $Q$ and $M$
for radio galaxies than for quasars in concordance with the faster decrease
of radio galaxy linear sizes, reported before (CZ). Furthermore, quasars,
contrary to the radio galaxies, show rather weak dependence on radio luminosity
(small $\beta $ values) in common for all the discussed geometrical parameters.
The differences between parameters found for radio galaxies and quasars are
statistically significant (at least at 95\% of confidence level) and seem to
contradict the simple unification scenario based entirely on the viewing angle.

We should emphasize that the above calculations were carried out using
all median values of $Q$, $M$, $L$ parameters estimated for our radio galaxies
and quasars, thus for samples not matched in redshift or luminosity. There are
to few median points to estimate the model parameters $\beta$ and $n$ at
satisfactory level of significance for radio galaxies and quasars restricted
to the matched regions in $P$-$z$ space. However, we have not observed any
abrupt change in the results when some median points were excluded from the
analysis. For example, we performed additional calculation to examine the
influence of high redshift quasars ($z>1.4$) on the results presented in
Table~\ref{QML}. For that purpose all quasars from four the highest-redshift
bins (see Fig.~\ref{FigPz}) were removed from the sample and whole fitting
procedure was repeated. The numbers we obtained

\[[-0.00\pm0.02;~ -0.15\pm0.12]: ~~\beta ~~{\rm and}~~ n ~~ {\rm for}~~ Q;\]

\[[+0.01\pm0.05;~ +0.48\pm0.53]: ~~\beta ~~{\rm and}~~ n ~~ {\rm for}~~ M;\]

\[[-0.01\pm0.06;~ -1.93\pm0.43]: ~~\beta ~~{\rm and}~~ n ~~ {\rm for}~~ L;\]

\noindent
indicate the similar to all quasars behavior of $Q$, $M$ and~$L$ parameters.
Of course, the errors are increased due to the~smaller data
set, but the estimated values of $\beta$ and $n$ still remain in 95\% of
confidence intervals presented in Table~\ref{QML}.
\begin{table*}[t]
 \begin{centering}
  \caption{The best-fit kinematical model parameters for radio galaxies
   and quasars for different range of redshift and opening angle of obscuring
   torus $\psi$. The probabilities (in percent) measure the goodness of fits.
   In the last section the upper rows refer to radio galaxies and the lower
   ones to quasars.}
\label{3-pkt}
\begin{array}[t]{l@{\hspace{5mm}}c@{\hspace{5mm}}rr@{\hspace{5mm}}rr@{\hspace
{5mm}}rrr@{\hspace{5mm}}rr@{\hspace{5mm}}rr@{\hspace{5mm}}rr}
\hline
\noalign{\smallskip}
   &  & \multicolumn{6}{c}{0.2<z<0.7} && \multicolumn{6}{c}{0.7<z<1.4} \\
\noalign{\smallskip}
\cline{3-8} \cline{10-15}
\noalign{\smallskip}
   &  &  \multicolumn{6}{c}{\psi}     && \multicolumn{6}{c}{\psi} \\
\noalign{\smallskip}
\cline{3-8} \cline{10-15}
\noalign{\smallskip}
   &  & \multicolumn{2}{c}{35^\circ} & \multicolumn{2}{c}{45^\circ} &
   \multicolumn{2}{c}{60^\circ} && \multicolumn{2}{c}{35^\circ} &
   \multicolumn{2}{c}{45^\circ} & \multicolumn{2}{c}{60^\circ}\\
\noalign{\smallskip}
\cline{3-8} \cline{10-15}
\\[1mm]

RG & \mu_{max} & 27^\circ& 36\%& 29^\circ& 28\%& 29^\circ& 20\%
   && 30^\circ& 82\%& 31^\circ& 81\%& 31^\circ& 82\%\\
   & k         &4.3& 95\%&4.2& 95\%&4.1& 93\%
   && 2.6&100\%&2.6&100\%&2.6&100\%\\
   & \sigma_T [10^7]  &3.5& 98\%&3.3& 98\%&3.1& 97\%
   && 2.7& 99\%&2.4&100\%&2.2&100\%\\
\\
QSO & \mu_{max} & 13^\circ&100\%& 15^\circ&100\%& 18^\circ&100\%
   && 16^\circ&100\%& 19^\circ&100\%& 23^\circ&100\%\\
   & k          &6.2& 93\%&6.0& 94\%&5.8& 94\%
   && 4.3&100\%&4.2&100\%&4.1&100\%\\
   & \sigma_T [10^7]  &7.3&100\%&5.8&100\%&5.0&100\%
   && 4.5&100\%&3.7& 99\%&3.0& 99\%\\
\\
RG+QSO & \mu_{max} & 19^\circ& ^{\textstyle 2\%}_{\textstyle 6\%}
   & 23^\circ&^{\textstyle 7\%}_{\textstyle 3\%}& 25^\circ&^{\textstyle
   12\%}_{\textstyle 16\%} && 22^\circ&^{\textstyle 22\%}_{\textstyle 27\%}
   & 25^\circ&^{\textstyle 42\%}_{\textstyle 49\%}& 28^\circ&^{\textstyle
   73\%}_{\textstyle 70\%}\\[2mm] & k & 5.1& ^{\textstyle 10\%}_{\textstyle
   11\%}& 5.0&^{\textstyle 13\%}_{\textstyle 13\%}& 4.8&^{\textstyle
   19\%}_{\textstyle 20\%}&& 3.4&^{\textstyle 26\%}_{\textstyle 37\%}
   & 3.3&^{\textstyle 31\%}_{\textstyle 28\%}& 3.2&^{\textstyle
   39\%}_{\textstyle 29\%}\\[2mm] & \sigma_T [10^7] & 5.0& ^{\textstyle
   0\%}_{\textstyle 3\%}& 4.3&^{\textstyle 4\%}_{\textstyle 5\%}& 3.8
   &^{\textstyle 16\%}_{\textstyle 17\%}&& 3.4&^{\textstyle 43\%}_{\textstyle
   40\%}& 2.9&^{\textstyle 63\%}_{\textstyle 50\%}& 2.6&^{\textstyle
   74\%}_{\textstyle 78\%}\\[2mm]
\hline
 \end{array}
 \end{centering}
\end{table*}

\begin{figure}[t]
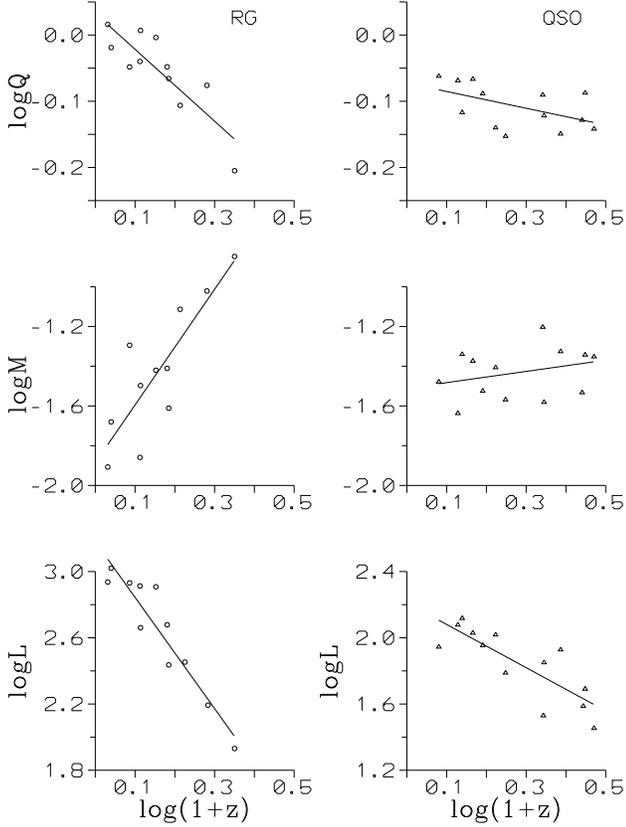

       \psbox{fig2a.psu}
      \caption{Estimated cosmological evolution for geometrical parameters
      $Q$, $M$, $L$ of the radio galaxy ($\circ$) and quasar ($\triangle$)
      structures}
   \label{FigQML-Z}
   \end{figure}

\begin{figure}[t]
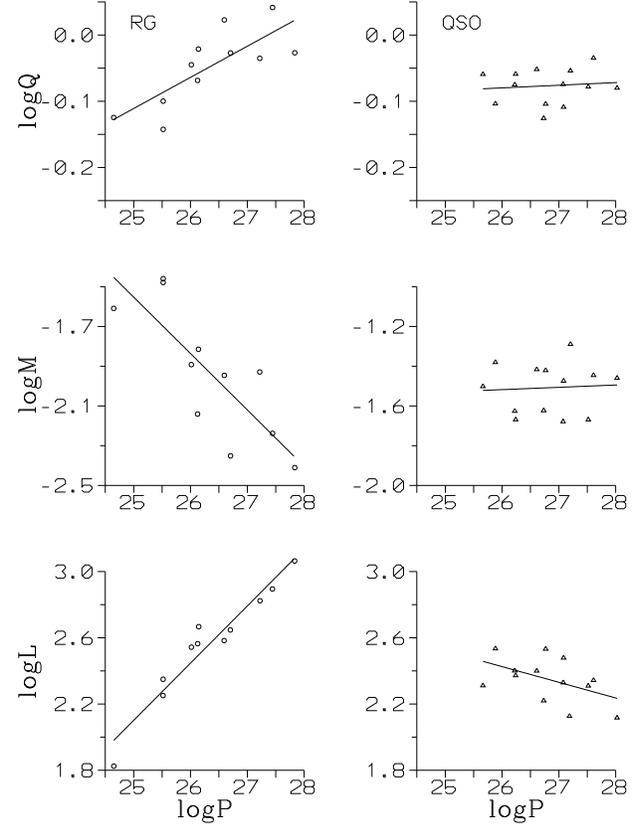

       \psbox{fig3a.psu}
       \caption{Estimated dependency of radio luminosity on geometrical
      parameters $Q$, $M$, $L$ of the radio galaxy ($\circ$) and quasar
      ($\triangle$) structures}
   \label{FigQML-P}
   \end{figure}

The possible causes underlying for the evolution of geometrical parameters
seems to be similar for radio galaxies and quasars. For both types of objects
with increasing redshift there is a fast decrease in their sizes, increase in
bending and slower increase in arm asymmetry (see Fig.~\ref{FigQML-Z}).
However, this epoch dependency is far stronger for radio galaxies than for
quasars which resemble analogous tendency in correlation with luminosity. The
dependence of the structure parameters on radio luminosity is presented on
Fig.~\ref{FigQML-P} where differences between galaxies and quasars are even
more visible.

\begin{figure}[t]
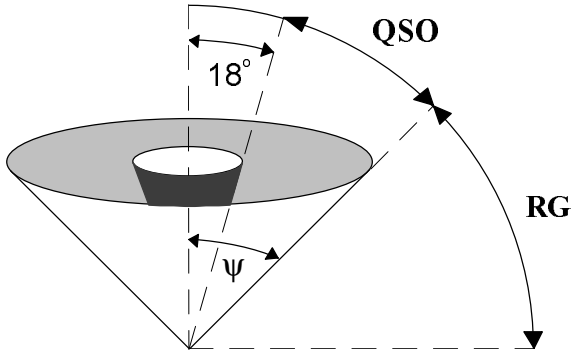

       \psxoffset=4cm\psyoffset=1cm
       \psbox{fig4a.psu}
      \caption{Visibility conditions for powerful radio
      galaxies and extended quasars tested by the kinematical model
      with the opening angle of obscuring torus $\psi$}
      \label{FigUNIF}
   \end{figure}

The comparison of observed structures of radio galaxies and quasars can be
also performed by another, distinct method. Having, for both analysed samples,
the observed distributions of $Q$, $M$ and $L$ we looked for the best
parameters
of the three-point kinematical model that can reproduce the observational data,
and explored ranges of values, and the sensitivities of the results to them.
According to the model, two plasmons, simultaneously and non-collineary ejected
from the core at two opposite sides, propagate through the external medium with
constant but different velocities. The three-point representation of a radio
source (the core plus two plasmons) is projected onto the sky, taking into
account the Doppler effect, thus producing an observed structure for which the
$Q$, $M$, $L$ parameters can be computed. The detailed description of the model
and the exact simulation procedure for achieving the model distribution of $Q$,
$M$ and $L$ was presented in our earlier paper (ZC). However, in order to make
the procedure consistent with the unification scheme, the line of sight angle
$\theta$, measured from the radio axis, was chosen at random from the
probability density $P(\theta) \propto$ sin$(\theta)$ within the viewing angle
interval [$18^{\circ}, \psi$] for quasars and [$\psi, 90^{\circ}$] for radio
galaxies (Fig.~\ref{FigUNIF}).

As quasars in our sample are triple sources, larger than $10$ kpc, we
eliminated from simulation objects seen almost along the radio axis, inside
the cone of $18^{\circ}$. This number follows from the comparison of quasars
counts in our sample and the 3CR complete data set.

The procedure for searching the model parameters was used separately for
different subsamples of radio galaxies and quasars constructed from our data
set. We distinguished quasars and radio galaxies in two redshift
intervals for which both types of objects are observed. The first spans a range
$0.2<z<0.7$ and the second $0.7<z<1.4$ (in Fig.~\ref{FigPz} they fill together
the $P$-$z$ space between plotted dividing lines).
This division gives a reasonable numbers of objects and does not join rather
different sources in one class (e.g. there is a deficiency of quasars in our
neighbour and scarcity of far away triple radio galaxies). The model
parameters needed to obtain the $Q$, $M$ and $L$ values of the modeled source
were as follows:
   \begin{enumerate}
      \item
$\mu_{max}~ -$ the largest intrinsic misalignment angle; a single intrinsic
misalignment angle was taken from the uniform distribution $[0, \mu_{max}]$
   \item
$k=\overline{v}/\sigma_{v}$;  the mean velocity $\overline{v}$ and the
standard deviation $\sigma_{v}$ describe the Guassian distribution from which
the expansion velocities of a model source were taken. As all simulations were
done for constant value of $\overline{v}=0.01c$ only $\sigma_{v}$ was treated
as a free parameter
   \item
$\sigma_{T}~ -$ the standard deviation of the Gaussian distribution
$(\overline{T}, \sigma_{T})$ from which the internal age $T$ was selected.
$\overline{T}=10^{7}$ years was accepted as a mean age for the all discussed
cases.
   \end{enumerate}
Once a set of model parameters was adopted, the $Q$, $M$ and $L$ model
distributions were obtained in the Monte Carlo simulation comprising
10000 generation. We then varied the model parameters to determine the fits,
based on the Kolmogorov test of consistency, that best reproduced the observed
$Q$, $M$ and $L$ distributions. It was done independently for three different
values of the viewing cone angle $\psi=35^{\circ}, 45^{\circ}$ and
$60^{\circ}$.

The results of the simulations are presented in Table~\ref{3-pkt}, which gives
the best fitted values of $\mu_{max}$, $k$ and $\sigma_{T}$, together with
the probabilities (in percent) that measure the goodness of fits, separately
for quasars and radio galaxy subsamples, and for the case when the both
classes of sources were treated as a one category of objects, having~the
some model parameters, but seen at different range of viewing angle
( [$18^{\circ}, \psi$] for quasars and  [$\psi,90^{\circ}$] for galaxies).

The numbers in Table~\ref{3-pkt} indicate that:
   \begin{enumerate}
   \item
it is difficult to find, at a significant level of the goodness of fit,
the same model parameters for quasars and galaxies treated in accordance with
the unified scheme
   \item
the large probabilities resulting from all the fitting procedures\footnote[1]
{the about 100\% goodness of fit for the observed $M$
distribution of radio galaxies was also possible to obtain, however the
intrinsic misalignment angle had to be taken not from a uniform distribution
but a Gaussian one (see also ZC)} in the cases of the pure quasar and radio
galaxy subsamples show that the rather different physical conditions,
disclosed by the unlike model parameters, are responsible for the observed
differences between quasars and radio galaxies.
   \end{enumerate}

\section{Discussion}

Extragalactic radio sources are observed as (elongated) structures being
rather far from spherical symmetry, so knowledge of their orientation to
our line of sight is an essential part of understanding their intrinsic
structure and possible strong selection effects. Properly speaking, in the
simple unified scheme proposed by Barthel (1989) the differences between
quasars and radio galaxies are simply treated as a result of a strong
orientation effect.

Recently, the selection approach to the radio galaxy-quasar problem was
favoured by Gopal-Krishna and Kulkarni's papers (Gopal-Krishna and Kulkarni
1992, Gopal-Krishna et al. 1994), which based on the observed radio sizes and
number densities of extragalactic radio sources. However, in order to explain
the discrepancy in~counts between radio galaxies and quasars reported
by~Singal (1993a), Gopal-Krishna et al. (1994) had to incorporate intrinsic
misalignment between radio axis and visibility cone in a rather large range
of angles (with a~mean $30^{\circ}$). The significant role of misalignment
effects,~but related to overall bending of a radio struc\-ture, was already
mentioned in our first paper (ZC) and recently~underlined in the work of
Lister et al. (1994) where orientation modelling of radio source properties
was presented. Although in conclusions Lister et al. accepted to some extent
orientation as a major parameter, they warned against their evidence as a
proof that there are no intrinsic differences too. They supported the
unifi\-cation~mainly on the base that the radio galaxy struc\-ture~modelled
with the parameters estimated from quasar~data are consistent with observed
radio galaxy morphology.

Unfortunately, they have not presented any statistical measure of this
consistency. As it is indicated in our analysis, finding satisfactory fits of
kinematical model with the~same parameters for both radio galaxies and quasars
is difficult, hence the intrinsic structure of extragalactic radio sources
must play an important role in the distinction between these two classes of
objects. Of course, the line of sight angle should be taken into account but
rather as a sort of factor which influence the observed structures of
physically different sources. In 1991 McCarthy et al. showed that the
asymmetry of radio source structures must be physically related to the
condition of the galaxy environment, namely the presence of thermal
line-emitting gas. According to their study, the closer of the two lobes
always lies on the side of high surface brightness optical line emission.
Among nearby galaxies, there is also observed the correlation of asymmetry
with radio luminosity (the objects which are less powerful are also more
asymmetric and smaller in size, e.g. ZC) which most likely arises from the
interaction of expanding structures with the surrounding medium. In that
case, one possible interpretation of the results from the estimation of
kinematical model and the comparison of evolutionary trends of $Q$, $M$,
$L$ parameters for quasar and radio galaxy structures is slightly different
state of galactic environment associated with these objects, even if their
deep interiors are identical.

An example supporting this point of view is the comparison of liner
sizes of nearby ($z<0.7$) radio sources. In our data sample of complete 3CR
and GB/GB2 surveys, contrary to the unification scheme prediction, the linear
sizes of radio galaxies (the median $131^{+14}_{-16}$) are less at $10\%$ of
significance level in Student's test than quasars sizes (with the median
$174^{+19}_{-51}$). A similar tendency was spotted by Singal (1993b) in a
large data set and interpreted as an evidence against the unified-scheme model.
The higher linear sizes of nearby quasars cannot be attributed to their
possible higher redshift in comparison to radio galaxies since if
the cosmological evolution of linear sizes is homogeneous the nearby quasars
sizes would be even underestimated.

One can be suspicious about this contradiction owing to broad line radio
galaxies (BLRG) which are sometimes treated as close quasars. In our sample
6 such objects are known. Because their median size is $187^{+40}_{-77}$ they
are certainly not smaller than ordinary quasars and their exclusion from
radio galaxies would even raise disagreement with the unification.

One should
also keep in mind, that the discrepancy with simple unification scenario is
visible on higher statistical levels when wider redshift range and the
appropriate trends are discussed, which Singal (1993a) reported and which can
be seen from the numbers in Table~\ref{QML} which were calculated for objects
with all possible redshifts. That is probably one of the reasons why our
results differ from those reported in Barthel (1989) and McCarthy et al. (1991)
papers. Besides, due to the dependence of $Q$, $M$ and $L$ parameter not only
on redshift but also on luminosity the appropriate radio structure analysis
should be fully 3-dimensional, in $(L, ~Q ~or ~L)$-$z$-$P$ space, as in
Singal's and our approach and contrary to the study in the two papers mentioned
above.

The obtained results can also be interpreted in the framework of a
subpopulation of radio galaxies which owing to different physical and
morphological properties separate from other galaxies and should not be
unified with quasars. Large linear sizes and small fraction of nearby quasars
suggest that such contaminating group of radio galaxies should consist of
objects with low radio power and small linear sizes.
Among close radio sources, there is already known a group of low luminosity,
small, edge-darkened (FRI) structures, with very weak emission lines and
no detectable broad lines and featureless continuum - so called dull galaxies
(Antonucci 1993). They are probably analogous to weak liners in a group of
radio quiet sources and therefore cannot be unified by orientation effects with
broad line objects. In our subsample we have only radio galaxies with FRII
morphology but few low redshift ones have radio luminosity below the arbitrary
FRI-II break at $10^{25}$ W/Hz at 1.4 GHz. However, the performed additional
estimation of $P^{\beta } (1+z)^{n}$ model for our radio galaxies but without
those weakest objects ($P<10^{25}$) has not revealed any statistical change
in the description of linear sizes evolution, thus differences between radio
galaxies and quasars certainly remain above the FRI-II break.

An interesting possibility is also to suppose that also some low power radio
galaxies of FRII type are optically dull, so they do not posses the broad line
region (BLR) and hence do not participate in the unification with quasars. In
fact, such a hypothesis has been put forward (Antonucci 1994) to explain
Singal (1993a) results but detailed dull object properties, redshift or
luminosity ranges have not been specified yet. To test this hypothesis we
performed the estimation of $P^{\beta } (1+z)^{n}$ model once again but without
the weakest and intermediate power radio galaxies ($P<10^{26}$, roughly
corresponding to $z<0.5$). This time the description of galaxy linear sizes
was different. The model parameters were estimated at not satisfactory
confidence level that may be attributed either to low density of objects in
$P$-$z$ space or to higher diversity of morphological properties of objects. In
any case, the possibility of dull FRII galaxies cannot be excluded, and the
reported modelling set up the redshift limit for the contaminating group
of galaxies as large as about $z\approx 0.5$. The existence of this separate
class of objects is appealing not only because it accounts for the observed
evolution of relative number and linear size of quasars and radio galaxies,
which otherwise cannot be explained in the framework of simple unification
scheme. Supposing the absent-BLR objects have also more asymmetric and bent
structures than the other galaxies (that is in concordance with observed
growing
of the asymmetry with smaller structures) they would also explain an intrinsic
asymmetry and bending for nearby radio galaxies higher than for quasars (see
in Table \ref{3-pkt} parameters $\mu_{max}$ and $k$ for the first redshift
range). Further spectropolarimetric observations of the all nearby radio
galaxies should reveal whether the subgroup of FRII objects without BLR really
exists and hence whether this supplementation of simple Barthel's unification
scheme is correct.

However, there is still a problem in this framework with high redshift radio
galaxies, which nevertheless seem to be more intrinsically asymmetrical and
bent than quasars (see Table \ref{3-pkt}). That contradiction cannot arise
from possible higher luminosity of these radio galaxies in comparison with
quasars as one might judge from Figure \ref{FigPz}: if the correlation of
asymmetry with luminosity for nearby galaxies still holds for the distant ones
we can speculate that less powerful high-redshift radio galaxies would be even
more asymmetrical and bent than those included in our data set.

\section{Conclusions}

The main results of this paper can be summarized as follows
   \begin{enumerate}
      \item
The investigation of radio structures of quasars and radio galaxies shows
that the cosmological evolution of geometrical properties of these two AGN
types are different. As well as linear size,  the arm asymmetry and bending
evolve more strongly with epoch for radio galaxies and their dependence on
radio luminosity is also stronger for radio galaxies than for quasars.
   \item
The performed estimation of kinematical model shows that finding the same set
of model parameters which reproduce radio galaxy and quasar observed
structures according to the simple unification scheme is not possible on high
level of statistical significance. The much better fits are achieved when the
structures of these two kinds of sources are modelled individually.
   \end{enumerate}

These findings seem to contradict the pure unification scenario based
entirely on the viewing angle and may reveal a slightly different state of
environmental conditions established during the evolution of these objects.

The other attractive possibility explaining the considered data is to admit
the existence of subpopulation of moderate redshift FRII galaxies without BLR
and with slightly smaller and more asymmetric structures than the remaining
part of the observed radio galaxies.

\acknowledgements
We would like to thank S. Ry\'{s} for fruitful conversation and to the
anonymous referee for helpful comments which contributed to improving our
paper. This work was supported by the grant from Polish Committee for
Scientific Research (KBN), grant no. PB/0595/P3/94/06.


\begin{thebibliography}{}

\bibitem{} Antonucci R., 1993, ARA\&A {\bf 31}, 473

\bibitem{} Antonucci R., 1994, in Multi-Wavelength Continuum Emission of
           AGN, IAU Symp. 159, eds. T.J.-L. Courvoisier, A. Blecha, Kluwer
           Academic Publishers, Dordrecht, p301

\bibitem{} Barthel P.D., Miley G.K., Schilizzi R.T., Lonsdale C.J.,
           1988, A\&AS {\bf 73}, 515

\bibitem{} Barthel P.D., 1989, ApJ {\bf 336}, 606

\bibitem{} Becker R.H., White R.L., Edwards A.L., 1991, ApJS {\bf 75}, 1

\bibitem{} Chy\.{z}y K.T., Zi\c{e}ba S., 1993, A\&A {\bf 267}, L27 (CZ)

\bibitem{} Fanaroff B., Riley J.M., 1974, MNRAS {\bf 167}, 31P

\bibitem{} Gopal-Krishna, Kulkarni V.K., 1992, A\&A {\bf 257}, 11

\bibitem{} Gopal-Krishna, Kulkarni V.K. and Mangalam A.V., 1994, A\&A
{\bf 268}, 459

\bibitem{} Hintzen P., Ulvestad J., Owen F., 1983, AJ {\bf 88}, 709

\bibitem{} Lister M.L., Hutchings J.B., Gower A.C., 1994, ApJ {\bf 427}, 125

\bibitem{} Machalski J., Maslowski J., 1982, AJ {\bf 87}, 1132

\bibitem{} Macklin J.T., 1981, MNRAS {\bf 196}, 967

\bibitem{} McCarthy P.J., van Breugel W., Kapahi V.K., 1991, ApJ {\bf 478}, 371

\bibitem{} Miley G.K., Hartsuijker A.P., 1978, A\&AS {\bf 34}, 129

\bibitem{} Oort M.J.A., 1987, Thesis, University of Leiden

\bibitem{} Ry\'{s} S., 1994, A\&A {\bf 281}, 15

\bibitem{} Singal A.,K., 1993a MNRAS, {\bf 262}, L27

\bibitem{} Singal A.,K., 1993b MNRAS, {\bf 263}, 139

\bibitem{} Spinrad H., Djorgovski S., Marr J., Aquilar L., 1985 PASP, {\bf 97},
   932

\bibitem{} White R.L., Becker R.H., 1992, ApJS {\bf 79},331

\bibitem{} Zi\c{e}ba S., Chy\.{z}y K.T., 1991, A\&A {\bf 241},22 (ZC)

\end{thebibliography}
\end{document}